\documentclass[fleqn,twoside,fps]{article}
\usepackage{espcrc2,graphicx,latexsym,amssymb,amsmath}
\usepackage[figuresright]{rotating}

\title{A comparative study of overlap and staggered fermions in
  QCD$^*$
}
\author{S. D\"urr\address[DESYZ]{DESY Zeuthen, Platanenallee 6,
        D--15738 Zeuthen, Germany}, 
  Ch. Hoelbling\address[CPTM]{Centre
        de Physique Th\'eorique$^\ddag$, Case 907, CNRS Luminy, F--13288
        Marseille Cedex 9, France}$^{\dag}$,
  U. Wenger\address[NICDESY]{NIC/DESY Zeuthen, Platanenallee 6,
        D--15738 Zeuthen, Germany}
}

\begin{document}

\begin{abstract}
We perform a  comparative study of the infrared  properties of overlap
and staggered fermions  in QCD. We observe that  the infrared spectrum
of the  APE/HYP improved staggered Dirac operator  develops a four-fold
near-degeneracy  and is  in  quantitative agreement  with the  infrared
spectrum  of the overlap  operator. The  near-degeneracy allows  us to
identify the  zero modes of  the staggered operator and  we find
that the  number of  zero modes is in line  with the
topological index of the overlap operator.
\end{abstract}

\maketitle

\footnotetext[1]{Talk presented by U.W.}
\footnotetext[2]{Supported by EU grant HPMF-CT-2001-01468.}
\footnotetext[3]{Unit{\'e} Mixte de Recherche (UMR 6207) du CNRS et des
Universit{\'e}s Aix Marseille 1, Aix Marseille 2 et sud Toulon-Var,
affili{\'e}e {\`a} la FRUMAM.}

\section{INTRODUCTION}
In the continuum the eigenmodes of the massless Dirac operator $D$ are
described by the eigensystem $D \psi_\lambda = i \lambda \psi_\lambda, \lambda
\in {\rm R\hspace{-0.3cm}I\hspace{0.2cm}}$.  Since the operator enjoys chiral
symmetry, $\{D,\gamma_5\} = 0$, and is $\gamma_5$-hermitian, its spectrum is
guaranteed to be symmetric around zero. The zero modes have a definite
chirality, $\psi_{\lambda=0}^\dagger \gamma_5 \psi_{\lambda=0} = \pm 1$, and
are related to the topological charge of the background gauge field via the
Atiyah-Singer index theorem which provides an explicit link to the axial
$U_A(1)$-anomaly and hence guarantees that the fermions are sensitive to
topology. On the other hand, the low-lying non-zero modes of the Dirac
operator describe the low-energy, long-distance physics and are therefore
referred to as the infrared (IR) modes.

A naive discretisation of the fermionic Dirac operator describes 16 flavours
of fermions instead of one.  A 4-fold degeneracy can be removed by
distributing the spin degrees of freedom over a hypercube yielding the
staggered Dirac operator
\begin{equation}
D^{\text{st}}_{x,y}=\frac{1}{2}\sum_{\mu} \eta_\mu(x)
    (U_\mu(x)\delta_{x+\hat\mu,y} - U^\dagger_\mu(x-\hat \mu)
    \delta_{x-\hat \mu,y})\nonumber
\end{equation}
where $\eta_\mu(x)=(-1)^{\sum_{\nu<\mu}x_\nu}$. This operator enjoys a remnant
$U(1)$ chiral symmetry which protects the bare quark mass from additive
renorma\-lisation, but it still describes 4 degenerate flavours in the
continuum. These doubler fermions are light and mix with each other generating
a complicated particle spectrum on the lattice. Moreover, since the operator
generically has no zero modes, there is no (exact) index theorem and it is not
clear, a priori, whether the operator is sensitive to topology.

There are other conceptual issues: in order to generate one flavour of sea
quarks in a dynamical simulation, one usually employs the so-called
fourth-root trick, i.e., one weights configurations by $\text{det}
(D^{\text{st}})^{1/4}$ instead of $\text{det} (D^{\text{st}})$ which, however,
is hard to justify from a field theoretical point of view. Even if it turns
out that this is indeed a valid procedure, the question remains how one should
match this prescription in the valence quark sector. The crucial point in this
context is therefore whether the above discretisation of fermions provides a
consistent field theoretic framework or whether it is just a model of QCD,
albeit a good one.

One way of avoiding (at least some of) the doubling problem is to suppress the
flavour changing interactions, i.e., to reduce the mixing of the doubler
modes. This can be achieved by removing ultraviolet (UV) noise from the gauge
fields by smearing the gauge links in one way or another.  In the following we
refer to this procedure as UV filtering or UV improvement and here we only
note that any such procedure simply corresponds to a $O(a^2)$ redefinition of
the staggered fermion action.

Another way is to make the doublers heavy, so that they decouple from the
theory in the continuum limit. A particular realisation of this idea is
provided by the overlap operator \cite{Narayanan:1993sk}
\begin{equation}\label{eq:overlap}
D_{\text{ov}}=\rho \big(1+\gamma_5 
      \text{sign}\large(\gamma_5 D_\text{w}(-\rho)\large)\big)
\end{equation}
where $D_\text{w}$ is the Wilson Dirac operator. The advantages of this
operator are manifold: there exists an exact lattice version of the continuum
chiral symmetry which guarantees the absence of an additive renormalisation of
the quark mass and avoids mixing of operators in different chiral
representations. Furthermore, the operator possesses exact zero modes and
allows to define an index theorem on the lattice implying the correct
sensitivity of the operator to topology.

Here we report on our results from a comparative study of the spectra of the
(UV improved) staggered and overlap Dirac operator \cite{Durr:2004as}.  We
concentrate on the IR spectra of the staggered operator and compare it to the
corresponding ones of the overlap operator as a benchmark.  The question we
would like to answer is whether one can see a `continuum-like' behaviour in
the spectra of the staggered operator similar to what has been observed in the
Schwinger model \cite{Durr:2003xs}, i.e.~a four-fold near-degeneracy and
approximate zero modes.  Then we would like to know at what lattice spacing it
appears, if at all, and whether any deviations scale like $O(a^2)$-artefacts.
The final goal, however, is to eventually test whether
\begin{equation}
\text{det}(D_{\text{st}})^{1/4} \propto 
       \text{det}(D_{\text{ov}})+O(a^2).
\label{eq:det}
\end{equation}

\begin{figure}[h]
\begin{center}
\includegraphics[width=7.5cm]{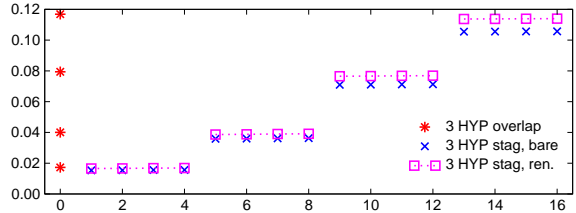} 
\end{center}
\vspace{-0.8cm}
\caption{{}Overlap and staggered eigenvalues
  for one of our quenched configurations at $a\simeq0.07$fm.}  
\vspace{-0.6cm}
\label{fig:degeneracy}
\end{figure}

\section{RESULTS}
We study the infrared spectra on four ensembles of quenched lattices with
lattice spacings ranging from $0.18$ to $0.07$fm matched to have a fixed
physical volume $V=(1.12\text{fm})^4$. We use standard smearing techniques and
we note that for the overlap operator we use $\rho=1.0$ throughout.  We find
that the overlap spectra are fairly stable under smearing for $a < 0.1$fm.
This indicates that the overlap operator separates well IR modes from UV
fluctuations at small enough lattice spacings.  The staggered spectra, on the
other hand, change dramatically under repeated smearings. In particular, we
find that without UV filtering there is no resemblance at all between the
staggered and the overlap spectra. With increased UV filtering the staggered
spectrum at large enough $\beta$ develops an (approximate) four-fold
degeneracy and zero modes start to separate out. This suggests that an
(approximate) index theorem holds for UV filtered staggered fermions. Moreover
we find a one-to-one correspondence between the quadruples of staggered
eigenvalues and the ones from the overlap operator. Occasionally, we find a
mismatch of topological charge between the index of the overlap and the
staggered Dirac operator. In particular we observe that whenever the overlap
operator has difficulties to decide what charge to assign to a configuration,
the staggered operator will have problems to produce a corresponding zero mode
-- instead, the four-fold degeneracy of the would-be zero mode remains rather
approximate. We assign this behaviour to the occurrence of dislocations with a
size of $O(1)$ lattice spacings which sometimes are detected as topological
charges and sometimes not.  It is clear, however, that even the overlap
operator suffers from this ambiguity (generically, this also happens when
changing the parameter $\rho$ in eq.(\ref{eq:overlap})) which suggests that
this is a genuine $O(a^2)$ effect.

In figure \ref{fig:degeneracy} we plot the chirally rotated overlap
eigenvalues most-left and those of the staggered operator ordered according to
increasing size for a generic configuration with $a=0.07$fm and UV filtering
level HYP3 \cite{Durr:2004as}.  After rescaling the staggered spectra with the
same factor for all configurations, they show a {\it quantitative} agreement
with the corresponding spectra of the overlap operator (up to mismatches in
the topological charge as discussed above).

\begin{figure}[t]
\begin{center}
\includegraphics[width=7.25cm]{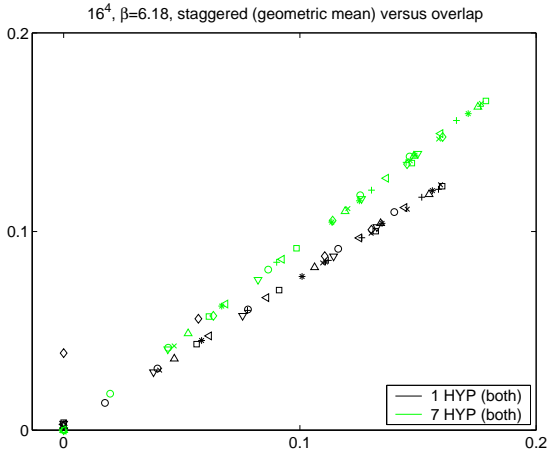}
\end{center}
\vspace{-1.2cm}
\caption{{}Staggered eigenvalues (geometric mean)
vs.~overlap eigenvalues at $a\simeq0.07$fm.
With only 1 HYP filtering, the overlap and staggered topological charges
disagree on one configuration, with 7 HYP steps there is no such mismatch. 
}
\vspace{-0.6cm}
\label{fig:correspondence}
\end{figure}

In figure \ref{fig:correspondence} we plot the geometric mean of the staggered
eigenvalues against the eigenvalues of the corresponding overlap operator for
filtering levels HYP1 and HYP7 on our quenched ensemble at $a\simeq0.07$fm. It
is evident that there is a linear correspondence between the UV filtered
overlap eigenvalues and the staggered eigenvalue quadruples at small
accessible couplings and with only moderate UV filtering. As emphasised
before, the correspondence of the (would-be) zero modes suggests an index
theorem for improved staggered fermions up to $O(a^2)$ artefacts.  The
one-to-one correspondence also implies that improved staggered fermions show
the same sensitivity to topology as overlap fermions and fall into the same
universality class of random matrix theory \cite{Shuryak:1992pi}. Indeed, this
has recently been verified \cite{Follana:2004sz,Wong:2004ai}.

\section{CONCLUSIONS AND OUTLOOK} 
We have found a four-fold degeneracy in the IR spectra of improved staggered
Dirac operators up to $O(a^2)$ artefacts, hence providing evidence that the
rooting-procedure of the staggered determinant might be justified in the sense
of eq.~(\ref{eq:det}).  We found a quantitative correspondence between the IR
part of the improved staggered and overlap spectra in QCD on individual
configurations and our data are compatible with the assumption that the
remaining artefacts scale as $O(a^2)$ with moderate UV filtering and at
accessible couplings.  However, a number of conceptual issues within the
staggered fermion formulation of QCD remain. In particular it is not clear,
whether and how a local operator $D$ with
$\text{det}(D)=\text{det}(D_{\text{stag}})^{1/4}$ can be constructed with the
staggered degrees of freedom.

Finally, we note that our results could point the way towards unquenching the
overlap Dirac operator via a `hybrid HMC' algorithm: one could use an improved
staggered Dirac operator along the molecular dynamics trajectory and
accept/reject the final configuration at the end of the trajectory with an
overlap Dirac operator. The hope is that in this way a correct dynamical
overlap ensemble could be generated essentially at the cost of a dynamical
staggered fermion simulation.

\end{document}